\documentclass[useAMS,usenatbib,fleqn]{mn2e}
\usepackage{listings}
\usepackage{amsmath}
\usepackage{natbib}
\usepackage{graphicx}
\usepackage{color} 
\usepackage{rotating}
\usepackage{nicefrac}
\usepackage{txfonts}
\usepackage{subfigure}
\usepackage{trivfloat}
\trivfloat{listfloat}
\usepackage{longtable}
\usepackage[percent]{overpic}
\usepackage{xspace}
\usepackage{url}
\usepackage{mathrsfs}
\usepackage{amssymb}
\usepackage{mathtools} 
\usepackage{mathtools, cuted}

\usepackage{enumitem}
\usepackage[percent]{overpic}
\usepackage{bbm}

\usepackage{epsfig}
\usepackage{amsmath}
\usepackage{rotating}
\usepackage{natbib}
\usepackage{multirow}
\usepackage[noend]{algpseudocode} 
\usepackage{flushend}

\newcommand{\ias}{{\sc \tt IAS15}\xspace}
\newcommand{\sabs}[1]{\ensuremath { \raisebox{0.3pt}{$\wr$} \mbox{{$#1$}} \raisebox{0.3pt}{$\wr$} } }

\lstdefinestyle{customc}{
  belowcaptionskip=1\baselineskip,
  language=C,
  showstringspaces=false,
  basicstyle=\footnotesize\ttfamily,
}

\newcommand\varr{\ensuremath{\mathbbm{r}}}
\renewcommand\varv{\ensuremath{\mathbbm{v}}}
\newcommand\varm{\ensuremath{\mathbbm{m}}}

\usepackage{array}
\usepackage{appendix}
\usepackage{comment}

\def \aap{A\&A}

\def \apjl{ApJ}
\def \apjs{ApJS}

\newcommand{\reb}{{\sc \tt REBOUND}\xspace}

\def\gsim{\;\rlap{\lower 2.5pt
 \hbox{$\sim$}}\raise 1.5pt\hbox{$>$}\;}
\def\lsim{\;\rlap{\lower 2.5pt
   \hbox{$\sim$}}\raise 1.5pt\hbox{$<$}\;}

\title{Second-order variational equations for $N$-body simulations}
\date{Submitted: 14 February 2016, Accepted: 10 March 2016.}

\author[Hanno Rein, Daniel Tamayo]{ 
Hanno Rein$^{1,2}$, Daniel Tamayo$^{1,3,4}$\\ 
$^1$ Department of Physical and Environmental Sciences, University of Toronto at Scarborough, Toronto, Ontario M1C 1A4, Canada\\
$^2$ Department of Astronomy and Astrophysics, University of Toronto, Toronto, Ontario, M5S 3H4, Canada\\
$^3$ Canadian Institute for Theoretical Astrophysics, 60 St. George St, University of Toronto, Toronto, Ontario M5S 3H8, Canada\\
$^4$ Centre for Planetary Sciences Fellow
}

\vspace{0.5\baselineskip}

\begin{document}
\maketitle

\begin{abstract}
First-order variational equations are widely used in $N$-body simulations to study how nearby trajectories diverge from one another.
These allow for efficient and reliable determinations of chaos indicators such as the Maximal Lyapunov characteristic Exponent (MLE) and the Mean Exponential Growth factor of Nearby Orbits (MEGNO).

In this paper we lay out the theoretical framework to extend the idea of variational equations to higher order.
We explicitly derive the differential equations that govern the evolution of second-order variations in the $N$-body problem.
Going to second order opens the door to new applications, including optimization algorithms that require the first and second derivatives of the solution, like the classical Newton's method.
Typically, these methods have faster convergence rates than derivative-free methods.
Derivatives are also required for Riemann manifold Langevin and Hamiltonian Monte Carlo methods which provide significantly shorter correlation times than standard methods.
Such improved optimization methods can be applied to anything from radial-velocity/transit-timing-variation fitting to spacecraft trajectory optimization to asteroid deflection.

We provide an implementation of first and second-order variational equations for the publicly available \reb integrator package. 
Our implementation allows the simultaneous integration of any number of first and second-order variational equations with the high-accuracy IAS15 integrator.
We also provide routines to generate consistent and accurate initial conditions without the need for finite differencing.
\end{abstract}


\section{Introduction}
\label{sec:intro}
Calculating the orbital motion of planets and predicting the position of planets in the night sky is one of astronomy's oldest reoccurring tasks. 
Today this is considered a solved problem, a simple application of Newtonian physics.
Typically the dynamical system is solved by numerically integrating forward in time using an $N$-body integrator.
Different techniques are available to do this very accurately over both short \citep[see e.g.][]{ReinSpiegel2015} and long \citep[see e.g.][]{WisdomHolman1991,ReinTamayo2015} timescales.

But often the solutions of even very simple dynamical systems are complex, in some cases exhibiting chaos.
This means that small perturbations in the initial conditions lead to exponentially diverging solutions at late times.
The solar system is one such chaotic dynamical system \citep{Roy1988,SussmanWisdome1988,LaskarGastineau2009}.
One way to characterize chaotic systems is to numerically determine the Maximal Lyapunov characteristic Exponent (MLE), which measures the rate of exponential divergence between neighbouring trajectories in phase space. 

Calculating how particle trajectories vary with respect to their initial conditions is therefore an important numerical task in modern celestial mechanics.
It is also immediately relevant to orbital fitting and optimization.
For example, when fitting an $N$-body simulation of a planetary system to data, one might want to calculate the derivative of the $\chi^2$ value with respect to a planet's initial orbital eccentricity.  

The MLE, or more generally any derivative with respect to initial conditions, can be calculated by running a separate $N$-body simulation with \emph{shadow particles}, where the initial conditions of one or more particles are slightly perturbed. 
Measuring how fast the distance in phase space of the shadow particles with respect to their unperturbed counterparts grows then yields the MLE \citep{Benettin1976}.

However, it is well known that there are problems associated with this shadow-particle method \citep{Tancredi2001}.
On the one hand, we want to start the shadow particles close so that we obtain a \emph{local} measure of the divergence of trajectories, and so that as the paths begin to drift apart, there are several decades over which to characterize the rate of divergence.
On the other hand, the closer we put the shadow particles, the more digits we lose to numerical roundoff error.
One workaround is to periodically rescale the separation vectors to keep shadow particles nearby their real counterparts \citep[see, e.g., Sec. 9.3.4 of][]{solarsystemdynamics}.
However, we show in Sec.~\ref{sec:finitedifference} that the use of shadow particles requires problem-dependent fine-tuning and that the problem is exacerbated when computing higher-order derivatives.

Luckily, instead of integrating a separate simulation of shadow particles, one can also use variational equations to measure divergences.
Rather than differencing two nearly equal trajectories, one instead derives a new linearized dynamical system that directly evolves the small distance between two initially offset particles.
These variational equations are scale-free and circumvent the numerical pitfalls associated with the shadow-particle method \citep{Tancredi2001}.

First-order variational equations have been widely discussed and applied in the literature \citep[e.g.,][]{MikkolaInnanen1999, Tancredi2001, Cincotta2003}.
In this paper we derive second-order variational equations for the $N$-body problem for the first time.
These provide the second derivatives of the solution with respect to the initial conditions.
Although mathematically straightforward to calculate, the number of terms and therefore the complexity rises significantly.
As we will see below, some terms involve 7 different (summation) indices.

Our work opens up many new opportunities for a variety of applications.
Perhaps most importantly, it is now straightforward to implement derivative-based optimization methods.
While the first derivatives provide a gradient that yields the {\it direction} towards a local minimum on a $\chi^2$ landscape, the second derivatives provide the {\it scale} for how far one should move to reach the minimum.
This can, among other things, dramatically improve fitting algorithms for radial velocity and transit planet searches, posterior estimation using Markov Chain Monte Carlo (MCMC) methods and even spacecraft trajectory optimization.

We begin in Sec.~\ref{sec:derivation} with a formal introduction to variational equations that generalizes to higher order.
In Sec.~\ref{sec:varnbody} we specialize to the case we are interested in, the $N$-body problem.
For completeness, in Sec.~\ref{sec:derivation1} we rederive the first-order variational equations for the $N$-body problem. 
We then go one step further in Sec.~\ref{sec:derivation2} and derive the second-order variational equations.

We have implemented the second-order variational equations within the \reb package and make them freely available.
\reb is a very modular $N$-body package written in \texttt{C99} and comes with an optional \texttt{python} interface.
We have abstracted the complexity of higher order variations significantly, and summarize our adopted syntax in Sec.~\ref{sec:implementation}.
Obtaining consistent initial conditions for variational equations in terms of Keplerian orbital elements without relying on finite difference is non-trivial.
We have therefore also implemented several convenience methods for this purpose.

In Sec.~\ref{sec:tests} we demonstrate how second-order variational equations and Newton's method can be used to fit observational data to a dynamical $N$-body model.
Finally, we compare variational and finite-difference methods in Sec.~\ref{sec:finitedifference}, and conclude in Sec.~\ref{sec:conclusions} by outlining the next steps in using higher order variational equations efficiently in optimization problems and MCMC methods.

\section{Derivation of differential equations}
\label{sec:derivation}
\subsection{Variational Equations} \label{sec:vareq}
In this section, we define what we mean by variational equations and introduce our notation. 
We follow the work of \cite{MoralesRuiz2007} and start with an analytic differential equation of the form
\begin{eqnarray}
\dot x = X(x).\label{eq:de}
\end{eqnarray}
In the case that we are interested in later, $x \in \mathbb{R}^{6N}$ encodes the 3 position and 3 velocity coordinates for each particle.  
$X$ is then a vector field on $\mathbb{R}^{6N}$. 
The dot represents a time derivative.
Given a suitable set of initial conditions $x_0$, an $N$-body simulation allows us to calculate (or at least approximate) the solution to Eq~\ref{eq:de}. 
We denote this solution $\phi(x_0,t)$, a $6N$ dimensional vector that depends on the initial conditions  $x_0\equiv x(0)$ and time $t$.

Our goal is to estimate the solution vector $\phi$ for different initial conditions, i.e. we want to approximate $\phi(y_0,t)$. 
One way to do that is to simply solve the differential equation in Eq.~\ref{eq:de} with the new initial conditions $y_0$. 
However, depending on the problem, finding the new solution with an $N$-body integration can be either very inefficient or inaccurate\footnote{In particular, it might be inaccurate if we are interested in the difference of the two solutions. See the discussion in Sec.~\ref{sec:intro} about shadow particles.}. 

Thus, we are looking for a better way to estimate solutions for the initial conditions $y_0$ in a neighbourhood of $x_0$.
The approach we consider here uses the fact that one can expand $\phi(y_0,t)$ around a reference solution $\phi(x_0,t)$ in a power series.
For simplicity, we first consider the case of varying a single scalar parameter $\alpha$ on which the initial conditions depend, $y_0(\alpha)$.
If $\alpha=0$, then $y_0(\alpha) = x_0$.
This could correspond to varying a Cartesian component of $x_0$, or a parameter that mixes Cartesian components, such as a planet's orbital eccentricity.
Then each component of $\phi(y_0,t)$ can be expanded around the reference solution as a power series in $\alpha$,
\begin{eqnarray}
\phi(y_0, t) = \sum_{m\geq 0} \frac{1}{m!}\; \phi^{(m)}\; \alpha^m, \label{eq:powerseries}
\end{eqnarray}
In the above equation, $\phi^{(0)} \equiv \phi(x_0,t)$, i.e. the reference solution, and 
\begin{align}
\phi^{(m)} \equiv \left.\frac{\partial^m \phi(y_0(\alpha), t)}{\partial \alpha^m}\right|_{\alpha=0}, \label{eq:phi-m}
\end{align}
i.e. a vector of the $m$-th derivative of each of the reference solution's components with respect to the parameter $\alpha$.
For sufficiently small $\alpha$ and $t$, this approximation is accurate even if we terminate the series at a finite $m=m_{\rm max}$ . 
The precise domain on which the solution can be trusted depends on the system and the initial conditions.
For example, in chaotic dynamical systems, $\phi^{(1)}$ might grow exponentially fast, limiting the domain to relatively short times or small $\alpha$.

In conclusion, if one can obtain the $\phi^{(m)}$, one can approximate all nearby solutions of $\phi(x_0, t)$.
Each $\phi^{(m)}$ is a function of time and must be numerically integrated.  
We therefore seek their governing differential equations.

We henceforth denote the reference solution $\phi(x_0, t)$ simply as~$\phi$.
The solution $\phi$, by definition, satisfies the original Eq.~\ref{eq:de}, in other words $ \dot{\phi} = X\left(\phi\right)$. 
We now take the derivative of this equation with respect $\alpha$, the parameter we are varying,
\begin{eqnarray}
\frac{\partial}{\partial t}\; \frac{\partial \phi}{\partial \alpha}  =
\frac{\partial X(\phi)}{\partial \alpha} =
\sum_b \frac{\partial X(\phi)}{\partial \phi_b} 
\frac{\partial \phi_b}{\partial \alpha},
\end{eqnarray}
where we changed the order of the derivatives and made use of the chain rule.
The summation index $b$ runs over all $6N$ elements of the vector $\phi$.
The derivative of $\phi$ with respect to $\alpha$ is the $\phi^{(1)}$ we seek for use in Eq.~\ref{eq:powerseries}.
Let us define the $6N$ by $6N$ matrix $X^{(1)}$ with components
\begin{eqnarray}
X_{ab}^{(1)}(\phi) \equiv \frac{\partial X_a(\phi)}{\partial \phi_b}. \label{eq:x1}
\end{eqnarray}
Using the matrix $X^{(1)}$ we then arrive at a compact set of differential equations for the vector $\phi^{(1)}$:
\begin{eqnarray}
\dot{\phi}^{(1)} = X^{(1)}(\phi)\; \phi^{(1)}. \label{eq:ve1}
\end{eqnarray}
This equation is the first-order variational equation.
We will later calculate the components of the matrix $X^{(1)}$ explicitly.
We then solve for the vector $\phi^{(1)}$ by integrating the differential equation numerically.
Note that $X^{(1)}$ depends on the time-dependent reference solution $\phi$ but not on $\phi^{(1)}$. 
It is a linear operator acting on $\phi^{(1)}$.

Repeating the steps above but differentiating Eq.~\ref{eq:de} twice instead of once, we can write down the differential equation for  $\phi^{(2)}$.
Because we apply the chain rule in the process, one finds that the time derivative of the second-order variations $\phi^{(2)}$ depends not only on $\phi^{(2)}$ but also on $\phi^{(1)}$.
Explicitly, the differential equation after two derivatives becomes
\begin{align}
&\frac{\partial}{\partial t}\; \frac{\partial^2 \phi}{\partial \alpha^2}  =
\frac{\partial^2 X(\phi)}{\partial \alpha^2} 
=
\sum_{b,c}
\frac{\partial^2 X(\phi)}{\partial \phi_b\partial \phi_c} 
\frac{\partial \phi_b}{\partial \alpha} 
\frac{\partial \phi_c}{\partial \alpha}
 + \sum_{b}
\frac{\partial X(\phi)}{\partial \phi_b} 
\frac{\partial^2 \phi_b}{\partial \alpha^2}.
\end{align}
Defining the $6N$ by $6N$ by $6N$ tensor $X^{(2)}$ with components 
\begin{eqnarray}
X_{abc}^{(2)}(\phi) \equiv \frac{\partial X_a(\phi)}{\partial \phi_b \partial \phi_c},
\end{eqnarray}
and using a short hand notation that suppresses the summation indices as well as the function arguments (we give explicit component forms for the general case in Sec.~\ref{sec:multi}), we have 
\begin{align}
\dot{\phi}^{(1)} &= X^{(1)}\;  \phi^{(1)}, \nonumber \\ 
\dot{\phi}^{(2)} &= X^{(1)}\;  \phi^{(2)} + X^{(2)}\;  \left[\phi^{(1)}\right]^2. \label{eq:ve2}
\end{align}
This set of equations is not linear anymore. 
But note that the linear term of the second line is the same as in the first line.

Higher order equations can be constructed in a straightforward way. 
Using the shorthand notation makes this particularly easy.
One can reintroduce the indices at the end of the calculation.
In this paper, we will only use variational equations up to second order \citep[see e.g.][for equations up to order 3]{MoralesRuiz2007}.

\subsection{Initial conditions}
To integrate a differential equation forward in time, one needs appropriate initial conditions.
To obtain the initial conditions for $\phi^{(1)}$ and $\phi^{(2)}$, one simply applies the chain rule to Eq.~\ref{eq:phi-m} and evaluates it at $t=0$,
\begin{align}
\phi^{(1)}(x_0,0) &= \left. \frac{\partial y_0(\alpha)}{\partial \alpha}, \label{eq:phi1ic} \right|_{\alpha=0}\\
\phi^{(2)}(x_0,0) &= \left. \frac{\partial^2 y_0(\alpha)}{\partial \alpha^2}\right|_{\alpha=0}. \label{eq:phi2ic}
\end{align}

In the case where the varied parameter $\alpha$ corresponds to a Cartesian component, choosing the initial conditions for $\phi^{(1)}$ and $\phi^{(2)}$  is straightforward.
If we assume the varied parameter has the coordinate index $b$, then the initial conditions for $\phi(y_0(\alpha),t)$ in component form are
\begin{align}
 \phi_a(y_0(\alpha),0) = \phi_a(x_0,0) + \alpha \delta_{ab}, 
\end{align}
where $\delta_{ab}$ is the Kronecker delta.
Thus
\begin{align}
\phi_a^{(1)}(x_0,0) = \delta_{ab}
\quad\quad\text{and} \quad\quad
\phi_a^{(2)}(x_0,0) = 0. \label{eq:initcart}
\end{align}

In practice, the function $y_0(\alpha)$ can be very complicated.
As an example, let us consider a planetary system with one planet of mass $m$ on an initially circular and coplanar orbit around a star with mass $M$.
The initial conditions of the planet might then be defined through the semi-major axis $a$ as
\begin{align}
    \begin{pmatrix}
    {r}_x\\
    {r}_y\\
    {r}_z
    \end{pmatrix}
    = 
    \begin{pmatrix}
    a\\
    0\\
    0
    \end{pmatrix},
    \quad
    \quad
    \begin{pmatrix}
    v_x\\
    v_y\\
    v_z
    \end{pmatrix}
    = 
    \begin{pmatrix}
    0\\
    \sqrt{G(m+M)/a}\\
    0
    \end{pmatrix}.
\end{align}
If we vary the initial semi-major axis by some length $\alpha$, then the initial conditions for the first-order variation are given by Eq.~\ref{eq:phi1ic}, in our case
\begin{align}
    \label{eq:init1st}
    \begin{pmatrix}
    \mathbbm{r}_x\\
    \mathbbm{r}_y\\
    \mathbbm{r}_z
    \end{pmatrix}^{(1)}
    = 
    \begin{pmatrix}
    1\\
    0\\
    0
    \end{pmatrix},
    \quad
    \quad
    \begin{pmatrix}
    \mathbbm{v}_x\\
    \mathbbm{v}_y\\
    \mathbbm{v}_z
    \end{pmatrix}^{(1)}
    &= 
    \begin{pmatrix}
    0\\
    -\frac12 \sqrt{G(m+M)/a^3}\\
    0
    \end{pmatrix}.
\end{align}
The initial conditions for the second-order variation are given by Eq.~\ref{eq:phi2ic}, which for the present case are
\begin{align}
    \label{eq:init2nd}
    \begin{pmatrix}
    \mathbbm{r}_x\\
    \mathbbm{r}_y\\
    \mathbbm{r}_z
    \end{pmatrix}^{(2)}
    = 
    \begin{pmatrix}
    0\\
    0\\
    0
    \end{pmatrix},
    \quad
    \quad
    \begin{pmatrix}
    \mathbbm{v}_x\\
    \mathbbm{v}_y\\
    \mathbbm{v}_z
    \end{pmatrix}^{(2)}
    &= 
    \begin{pmatrix}
    0\\
    \frac34 \sqrt{G(m+M)/a^5}\\
    0
    \end{pmatrix}.
\end{align}
The components of $\phi^{(1)}$ and $\phi^{(2)}$ that we calculated above correspond to the planet.
All components corresponding to the star are~0. 

The initialization can quickly get complicated.
Suppose we work in the centre-of-mass frame. 
Then the star's initial conditions will also depend on the semi-major axis of the planet.
Similarly, if we add an additional outer planet and work in Jacobi coordinates, the outer planet's initial conditions depend on the inner planet's orbital parameters.
For that reason we've implemented convenience functions for the initialization of orbits which we present later in Sec.~\ref{sec:syntax}. 

\subsection{Multiple sets of variational equations} \label{sec:multi}
The above derivation of variational equations can be straightforwardly generalized when varying multiple parameters.  
Consider varying the initial value of $N_{\rm par}$ separate parameters $\alpha_\xi$.
Here and in the rest of this paper Greek variables indicate to variations with respect to one parameter and will run over the interval $[0,N_{\rm par}-1]$.
We write all equations in this section in component form for direct comparison with our later results.

When varying several parameters, the coupled set of differential equations, Eq.~\ref{eq:ve2}, becomes
\begin{align}
\dot{\phi}_{a,\xi}^{(1)} &= \sum_{b} X_{ab}^{(1)}\;  \phi_{b,\xi}^{(1)}, \nonumber \\ 
\dot{\phi}_{a,\xi\eta}^{(2)} &= \sum_{b} X_{ab}^{(1)}\;  \phi_{b,\xi\eta}^{(2)} + \sum_{b,c} X_{abc}^{(2)}\; \phi_{b,\xi}^{(1)} \phi_{c,\eta}^{(1)}, \label{eq:ve2multi}
\end{align}
where
$\phi_{a,\xi}^{(1)} \equiv \frac{\partial \phi_a(x_0, t)}{\partial \alpha_\xi}$, and
$\phi_{a,\xi\eta}^{(2)} \equiv \frac{\partial^2 \phi_a(x_0, t)}{\partial \alpha_\xi \partial \alpha_\eta}$.
Therefore, when varying $N_{\rm par}$ parameters, there are $N_{\rm par}$ sets of first-order variational equations, one for each of the vectors $\phi_{\xi}^{(1)}$.
There are $N_{\rm par}^2$ sets of second-order variational equations (one for each of the vectors $\phi_{\xi\eta}^{(2)}$).
Each set of variational equations has $6N$ components.

Once numerically integrated, these variations can then be plugged into a multi-variate power series expansion analog to Eq.~\ref{eq:powerseries} to obtain trajectories for arbitrary nearby initial conditions.
Explicitly, to second order,  
\begin{eqnarray}
\phi_a(y_0, t) \approx \phi_a(x_0, t) + \sum_{\xi} \phi_{a,\xi}^{(1)} \alpha_\xi+ \frac12 \sum_{\xi,\eta}\phi_{a,\xi\eta}^{(2)} \alpha_\xi \alpha_\eta. \label{eq:powerseriesmulti}
\end{eqnarray}
Note that because derivatives commute we find that $\phi^{(2)}_{\xi \eta} = \phi^{(2)}_{\eta \xi}$. 
Thus the total number of differential equations we need to integrate for the second order variations can be reduced from $6\,N\,N^2_{\rm par}$ to $3\,N\,N_{\rm par} (N_{\rm par}+1)$.
This is in addition to the $6N$ differential equations for the reference simulation and $6\,N\,N_{\rm par}$ equations for the first order variations.

\subsection{Index convention}
As we saw above, the number of indices in second-order expressions is high. 
We therefore adopt a consistent index notation for the remainder of this paper.
Specifically, we will consider a dynamical system consisting of $N$ particles and use the indices $i,j, k $ and $l$ to label different particles.
These indices thus run from $0$ to $N-1$.
The indices $a,b,c$ and $d$ label coordinate axes.
Above, these indices ran over the $6N$ coordinates of the $N$-body system.
We will find below that for the $N$-body system, it is simpler to consider positions and velocities separately.
Therefore, in what follows $a,b,c$ and $d$ will run over the Cartesian $x$, $y$, and $z$ components only.
As before, we will also make use of Greek characters $\xi$ and $\eta$ to indicate different \emph{sets} of variational equations corresponding to different varied parameters (different \emph{variations}).
In the following sections we explicitly write summation symbols, i.e., we do not use a summation convention over repeating indices.

\section{Variational Equations for the $N$-body System}
\label{sec:varnbody}
Let us now derive the differential equations from above for a specific problem: the dynamical system of $N$ gravitationally interacting particles.
The differential equation for the $N$-body problem in vector notation is
\begin{eqnarray}
\ddot{\mathbf{r}}_i &=& -\sum_{j,\;j\neq i} \frac{G m_j{\pmb r}_{ij}}{r_{ij}^3}\label{eq:nbody}
\end{eqnarray}
where ${\pmb r}_{ij} = {\pmb r}_i - {\pmb r}_j$ and $r_{ij}$ is the norm of ${\pmb r}_{ij}$.
We can also write the equation in component form
\begin{eqnarray}
\dot{v}_{i,a} &=& -\sum_{j,\;j\neq i}\frac{ G m_j\,r_{ij,a}}{r_{ij}^3}. \label{eq:nbodycomp}
\end{eqnarray}
where ${r}_{ij,a} = {r}_{i,a} - {r}_{j,a}$ is the relative position between particles $i$ and $j$. 
This is a second-order differential equation. 
However, note that the first time-derivative of the position is just the velocity, $\dot{\pmb r} = {\pmb v}$.
The differential equation can thus easily be brought into the form of Eq.~\ref{eq:de} by introducing
\begin{align}
 \arraycolsep=2.4pt\def\arraystretch{1.}
x \equiv  
    \begin{pmatrix}
    r_{0,x}&
    r_{0,y}&
    \hdots&
    r_{N-1,y}&
    r_{N-1,z}&
    v_{0,x}&
    v_{0,y}&
    \hdots&
    v_{N-1,y}&
    v_{N-1,z}
    \end{pmatrix}^T
    \label{eq:x}
\end{align}
such that
\begin{align}
 \arraycolsep=2.4pt\def\arraystretch{1.}
\dot x =  
    \begin{pmatrix}
    v_{0,x}&
    v_{0,y}&
    \hdots&
    v_{N-1,y}&
    v_{N-1,z}&
    \ddot r_{0,x}&
    \ddot r_{0,y}&
    \hdots&
    \ddot r_{N-1,y}&
    \ddot r_{N-1,z}
    \end{pmatrix}^T.
\end{align}
We end up with a first-order differential equation with twice as many variables as Eq.~\ref{eq:nbody} ($6N$ compared to $3N$). 
This set of differential equations together with suitable initial conditions completely describes the $N$-body problem.

\subsection{First-order variational equations}
\label{sec:derivation1}
To derive the first-order variational equation for the $N$-body problem, we start by differentiating Eq.~\ref{eq:nbodycomp} with respect to $r_{j,b}$.
To be as explicit as possible we do this in component form.
We end up with an equation with four indices. 
Two of the indices run over coordinates, and two over particles,
\begin{align}
\frac{\partial \dot{v}_{i,a}}{\partial r_{j,b}} &= \frac{\partial}{\partial r_{j,b}}\left(
 -\sum_{k,\;k\neq i}\frac{ G m_k\,r_{ik,a}}{r_{ik}^3}. \nonumber
\right)
\\
&= -\sum_{k,\;k\neq i}\frac{Gm_k\,(\delta_{ij}-\delta_{kj})\delta_{ab} }{r_{ik}^3} 
 + \sum_{k,\;k\neq i}3 \frac{G m_k r_{ik,a}\,r_{ik,b}}{r_{ik}^5}(\delta_{ij}-\delta_{kj})\nonumber
 \\
&= 
\begin{cases} 
\sum_{k,\;k\neq i}\left( -\frac{Gm_k\,\delta_{ab} }{r_{ik}^3} 
+ 3 \frac{G m_k r_{ik,a}\,r_{ik,b}}{r_{ik}^5}\right) &\mbox{ if } i=j\\
+\frac{Gm_j\,\delta_{ab} }{r_{ij}^3} 
- 3 \frac{G m_j r_{ij,a}\,r_{ij,b}}{r_{ij}^5} & \mbox{ if } i\neq j
\end{cases}.
\end{align}
\noindent The above expression gives us the matrix elements of $X^{(1)}$ (Eq.~\ref{eq:x1}). 
Note that two indices $j$ and $b$ combined correspond to the row of that matrix, the other two ($i$ and $a$) correspond to the column of the matrix.

We also want to consider the influence of varying masses.
Note that one can think of the masses $m_i$ as part of the initial conditions.
However, we assume that the masses do not vary with time after the system has been initialized, thus $\dot m_i=0$.
We need the derivative of the force with respect to the mass:
\begin{eqnarray}
\frac{\partial \dot{v}_{i,a}}{\partial m_j} = 
\frac{\partial}{\partial m_j}\left(-\sum_{k,\;k\neq i}\frac{G\,m_k\,{r}_{ik,a}}{r_{ik}^3}\right)
=
\begin{cases} 
	-\frac{G\,{r}_{ij,a} }{r_{ij}^3}  & \mbox{if } j\neq i \\ 
	0 & \mbox{if } j = i. 
\end{cases} 
\end{eqnarray}

We can now write down the differential equation for the first-order variational equation using Eq.~\ref{eq:ve1}.
We could do this in terms of $\phi^{(1)}$ and its components, but choose to use two separate vectors for the variational position and velocity components (to be consistent with Eq.~\ref{eq:nbody}).
Variational quantities are denoted by double-striped symbols $\varr$, $\varv$ and $\varm$.
First-order variational quantities receive the superscript ${}^{(1)}$. 
Thus, one might write $\phi^{(1)}$ in terms of $\varr^{(1)}$ and $\varv^{(1)}$: 
\begin{eqnarray}
\phi^{(1)}= \begin{pmatrix}
    \varr_0^{(1)}&
    \hdots&
    \varr^{(1)}_{N-1}&
    \varv^{(1)}_0&
    \hdots&
    \varv^{(1)}_{N-1}
\end{pmatrix}^T
\end{eqnarray}
which should be compared to Eq.~\ref{eq:x}.
The units of $\varr^{(1)}$ and $\varv^{(1)}$ depend on the variation we are considering.
In general the units are not the same as those of $r$ and $v$ (see Eq.~\ref{eq:phi1ic}). 
We end up with the following set of equations for the components of $\dot \varv^{(1)}$ (corresponding to the second half of the components of $\dot \phi^{(1)}$):
\begin{align}
\dot{\varv} ^{(1)}_{i,a} &= 
\sum_j
\sum_b
\frac{\partial \dot{v}_{i,a}}{\partial r_{j,b}} 
 \varr^{(1)}_{j,b}
 +
\sum_j
\frac{\partial \dot{v}_{i,a}}{\partial m_j} 
\varm_j^{(1)},\nonumber
 \\
 &=
\sum_{j,\,j\neq i}
\sum_b
\frac{\partial \dot{v}_{i,a}}{\partial r_{j,b}} 
 \varr^{(1)}_{j,b},
+
\sum_b
\frac{\partial \dot{v}_{i,a}}{\partial r_{i,b}} 
 \varr^{(1)}_{i,b}
 +
\sum_j
\frac{\partial \dot{v}_{i,a}}{\partial m_j} 
\varm_j^{(1)},
\nonumber
 \\
 &=
\sum_{j,\,j\neq i}
\left[
\sum_b 
\left(
-\frac{Gm_j\,\delta_{ab} }{r_{ij}^3} 
+ 3 \frac{G m_j r_{ij,a}\,r_{ij,b}}{r_{ij}^5}
 \right)
  \varr^{(1)}_{ij,b} 
 -
\frac{G\,{r}_{ij,a} }{r_{ij}^3}
\varm_j^{(1)}
\right]
,
\end{align}
\noindent where $\varr^{(1)}_{ij,b} = \varr^{(1)}_{i,b} - \varr^{(1)}_{j,b}$. 
We can rewrite this in vector notation, which allows us to drop the indices $a$ and $b$:
\begin{eqnarray}
\dot{\pmb \varv} ^{(1)}_{i} &=& 
\sum_{j,\,j\neq i}
\left(
-\frac{Gm_j}{r_{ij}^3} 
  {\pmb \varr}^{(1)}_{ij} 
+ 3 \frac{G m_j {\pmb r}_{ij}}{r_{ij}^5}
  \left( {\pmb r}_{ij} \cdot {\pmb \varr}^{(1)}_{ij} \right)
	-\frac{G\,{\pmb r}_{ij} }{r_{ij}^3} \varm_j^{(1)} 
 \right)\label{eq:var}
\end{eqnarray}
The $\cdot$ represent the usual vector product.
The equations for the variational positions, $\varr^{(1)}$ (the first half of the components of $\phi^{(1)}$), are significantly easier to write down:
\begin{eqnarray}
\dot{\pmb \varr} ^{(1)}_{i} &=& {\pmb \varv}^{(1)}_{i}
\end{eqnarray}
As mentioned before, the masses are assumed to be constant throughout a simulation; thus, the variational equation for the mass coordinates are not dynamic:
\begin{eqnarray}
\dot{\varm} ^{(1)}_{i} &=& 0.
\end{eqnarray}
The solutions are trivial, $\varm_i(t) = \varm_i(0)$, and we therefore do not evolve the quantities $\varm_i$.

\begin{strip}
\subsection{Second-order variational equations}
\label{sec:derivation2}
We now derive the second-order variational equations.
As a warning to the reader: this will get messy.
We nevertheless present the calculation in full detail as it is easy to get confused with up to 7 indices in a single term.
This should ease derivations for alternate dynamical systems, for example if one want to include additional non-gravitational effects.
Conceptually, this is the same procedure as in the previous section, just to second order.
Because of the chain rule, we end up with significantly more terms.

We begin by calculating various second-order derivatives that we will need later.
The second-order derivative of the force with respect to the positions is
\begin{eqnarray} 
\frac{\partial^2 \dot{v}_{i,a}}{\partial r_{j,b} \partial r_{k,c}} &=& 
\frac{\partial}{\partial r_{k,c}}
\begin{cases} 
\sum_{l,\;l\neq i}\left( -\frac{Gm_l\,\delta_{ab} }{r_{il}^3} 
+ 3 \frac{G m_l r_{il,a}\,r_{il,b}}{r_{il}^5}\right) &\mbox{ if } i=j\\
+\frac{Gm_j\,\delta_{ab} }{r_{ij}^3} 
- 3 \frac{G m_j r_{ij,a}\,r_{ij,b}}{r_{ij}^5} & \mbox{ if } i\neq j
\end{cases}
\end{eqnarray}
We look at both cases individually.
The first case, $i=j$, gives

\begin{align} 
\frac{\partial^2 \dot{v}_{i,a}}{\partial r_{i,b} \partial r_{k,c}} 
&= 
\frac{\partial}{\partial r_{k,c}}
\sum_{l,\;l\neq i}\left( -\frac{Gm_l\,\delta_{ab} }{r_{il}^3} 
+ 3 \frac{G m_l r_{il,a}\,r_{il,b}}{r_{il}^5}\right) 
\nonumber
 \\
&= 
\sum_{l,\;l\neq i} 3\frac{Gm_l\,\delta_{ab} }{r_{il}^5} r_{il,c} (\delta_{ik} - \delta_{lk})  
 +
\sum_{l,\;l\neq i}
 3 \frac{G m_l \delta_{ac}}{r_{il}^5}  r_{il,b}(\delta_{ik}-\delta_{lk})
 +
\sum_{l,\;l\neq i}
 3 \frac{G m_l \delta_{bc}}{r_{il}^5} r_{il,a}(\delta_{ik}-\delta_{lk})
 -
\sum_{l,\;l\neq i}
 15 \frac{G m_l r_{il,a}\,r_{il,b}}{r_{il}^7} r_{il,c} (\delta_{ik}-\delta_{lk}) 
\nonumber
 \\
&=
\begin{cases}
\sum_{l,\;l\neq i}\left(
 3\frac{Gm_l\,\delta_{ab} r_{il,c} }{r_{il}^5} 
 +
 3 \frac{G m_l \delta_{ac}r_{il,b}}{r_{il}^5} 
 +
 3 \frac{G m_l \delta_{bc}r_{il,a}}{r_{il}^5} 
 -
 15 \frac{G m_l r_{il,a}\,r_{il,b}r_{il,c}}{r_{il}^7}   
 \right)
  & \mbox{ if } i=k\\
-\left(
 3\frac{Gm_k\,\delta_{ab} r_{ik,c}}{r_{ik}^5}  
 +
 3 \frac{G m_k \delta_{ac} r_{ik,b}}{r_{ik}^5}
 +
 3 \frac{G m_k \delta_{bc} r_{ik,a}}{r_{ik}^5}
 -
 15 \frac{G m_k r_{ik,a}\,r_{ik,b} r_{ik,c}  }{r_{ik}^7}
 \right)
 & \mbox{ if } i\neq k
\end{cases}
\end{align}

\noindent The second case, $i\neq j$, is similar but with the sign reversed and without the summation:

\begin{align} 
\frac{\partial^2 \dot{v}_{i,a}}{\partial r_{j,b} \partial r_{k,c}} 
&= - 
 3\frac{Gm_j\,\delta_{ab} }{r_{ij}^5} r_{ij,c} (\delta_{ik} - \delta_{jk})   
 -
 3 \frac{G m_j \delta_{ac}}{r_{ij}^5}  r_{ij,b}(\delta_{ik}-\delta_{jk})
 -
 3 \frac{G m_j \delta_{bc}}{r_{ij}^5} r_{ij,a}(\delta_{ik}-\delta_{jk})
+
 15 \frac{G m_j r_{ij,a}\,r_{ij,b}}{r_{ij}^7} r_{ij,c} (\delta_{ik}-\delta_{jk}) 
\nonumber
 \\
&=
\begin{cases}
 - 
 3\frac{Gm_j\,\delta_{ab} }{r_{ij}^5} r_{ij,c} 
 -
 3 \frac{G m_j \delta_{ac}}{r_{ij}^5}  r_{ij,b}
 -
 3 \frac{G m_j \delta_{bc}}{r_{ij}^5} r_{ij,a}
 +
 15 \frac{G m_j r_{ij,a}\,r_{ij,b}}{r_{ij}^7} r_{ij,c} 
 & \mbox{ if } i= k\\
 + 
 3\frac{Gm_j\,\delta_{ab} }{r_{ij}^5} r_{ij,c} 
 +
 3 \frac{G m_j \delta_{ac}}{r_{ij}^5}  r_{ij,b}
 +
 3 \frac{G m_j \delta_{bc}}{r_{ij}^5} r_{ij,a}
 -
 15 \frac{G m_j r_{ij,a}\,r_{ij,b}}{r_{ij}^7} r_{ij,c} 
 & \mbox{ if } i\neq k \mbox{ and } j=k\\
 0 & \mbox{ otherwise}
\end{cases}
\end{align}

\noindent We also need the derivatives with respect to the particles' masses.
Luckily, if we differentiate the force twice with respect to mass, we get zero:
\begin{eqnarray}
\frac{\partial^2 \dot{v}_{i,a}}{\partial m_j \partial m_k} &=& 0.
\end{eqnarray}
However, other second derivatives involving the mass are not zero:
\begin{eqnarray}
\frac{\partial^2 \dot{ v}_{i,a}}{\partial { r}_{k,b} \partial m_j} 
= \frac{\partial}{\partial r_{k,b} }\left(
\frac{\partial \dot{v}_{i,a}}{\partial m_k} \right) 
= \frac{\partial}{\partial r_{k,b} }
\begin{cases} 
	-\frac{G\,{r}_{ij,a} }{r_{ij}^3}  & \mbox{if } j\neq i \\ 
	0 & \mbox{if } j = i. 
\end{cases} 
\end{eqnarray}
Restricting ourselves to the $j\neq i$ case,
\begin{align}
\frac{\partial^2 \dot{ v}_{i,a}}{\partial { r}_{k,b} \partial m_j} 
= \frac{\partial}{\partial r_{k,b} }
	\left(-\frac{G\,{r}_{ij,a} }{r_{ij}^3}  \right) 
&=
	-\frac{G}{r_{ij}^3} \delta_{ab} (\delta_{ik}-\delta_{jk})
	+3\frac{G\,{r}_{ij,a} }{r_{ij}^5} r_{ij,b} (\delta_{ik}-\delta_{jk})
    =
\begin{cases} 
	+\frac{G}{r_{ij}^3} \delta_{ab} \delta_{jk}
	-3\frac{G\,{r}_{ij,a} }{r_{ij}^5} r_{ij,b} \delta_{jk}
	  & \mbox{if } k\neq i \\ 
	-\frac{G}{r_{ij}^3} \delta_{ab} 
	+3\frac{G\,{r}_{ij,a} }{r_{ij}^5} r_{ij,b} 
	 & \mbox{if } k = i, 
\end{cases} 
\end{align}
such that after putting all cases together we arrive at
\begin{align}
\frac{\partial^2 \dot{ v}_{i,a}}{\partial { r}_{k,b} \partial m_j} 
&=\begin{cases} 
	+\frac{G}{r_{ij}^3} \delta_{ab} 
	-3\frac{G\,{r}_{ij,a} }{r_{ij}^5} r_{ij,b}
	  & \mbox{if } j\neq i \mbox{ and } k = j \\ 
	-\frac{G}{r_{ij}^3} \delta_{ab} 
	+3\frac{G\,{r}_{ij,a} }{r_{ij}^5} r_{ij,b} 
	 & \mbox{if } j\neq i \mbox{ and } k = i\\ 
	 0
	 & \mbox{otherwise} . 
\end{cases} 
\end{align}

With the above expressions of the second order force derivatives, we can now construct the second-order variational equations.
At this point we introduce two more indices that describe the variation under consideration, $\xi$ and $\eta$.
They run over all the variations that we want to consider.
In vector notation Eq.~\ref{eq:ve2} can be expressed as

\begin{align}
\dot{\varv} ^{(2)}_{i,a,\eta\xi} &=
\sum_j
\sum_b
\frac{\partial \dot{v}_{i,a}}{\partial r_{j,b}} 
 \varr^{(2)}_{j,b,\eta\xi}
+ 
\sum_j
\sum_k
\sum_b
\sum_c
\frac{\partial^2 \dot{v}_{i,a}}{\partial r_{j,b} \partial r_{k,c}} 
{\varr}^{(1)}_{j,b,\xi}{\varr}^{(1)}_{k,c,\eta}
\nonumber
\\
&
+ 
\sum_j
\sum_k
\sum_b
\frac{\partial^2 \dot{ v}_{i,a}}{\partial { r}_{k,b} \partial m_j} 
{\varr}^{(1)}_{k,b,\xi}{\varm}^{(1)}_{j,\eta}
+ 
\sum_j
\sum_k
\sum_b
\frac{\partial^2 \dot{ v}_{i,a}}{\partial m_k\partial { r}_{j,b} } 
{\varm}^{(1)}_{k,\xi}
{\varr}^{(1)}_{j,b,\eta}
 +
\sum_j
\frac{\partial \dot{v}_{i,a}}{\partial m_j} 
\varm_{j,\eta\xi}^{(2)}
+
\underbrace{ 
\sum_j
\sum_k
\frac{\partial^2 \dot{ v}_{i,a}}{\partial { m}_{k} \partial m_j} 
{\varm}^{(1)}_{k,\xi}{\varm}^{(1)}_{j,\eta}
}_{=0}.
\end{align}

\noindent We replace the derivatives with what we calculated above.
The result is a rather long expression with 7 different (summation) indices:

\begin{align}
\dot{\varv} ^{(2)}_{i,a,\xi\eta} = &
\sum_{j,\,j\neq i}
\left(
-\frac{Gm_j\, \varr^{(2)}_{ij,a,\xi\eta} }{r_{ij}^3} 
+ 3 \frac{G m_j r_{ij,a}}{r_{ij}^5}
\sum_b 
\left(
r_{ij,b}
\cdot
  \varr^{(2)}_{ij,b,\xi\eta} 
 \right)
  -
	\frac{G\,{r}_{ij,a} }{r_{ij}^3} \varm_{j,\xi\eta}^{(2)} 
 \right)
\nonumber
\\
&
+ 
\sum_b
\sum_c
\sum_{l,\;l\neq i}\left(
 3\frac{Gm_l\,\delta_{ab} r_{il,c} }{r_{il}^5} 
 +
 3 \frac{G m_l \delta_{ac}r_{il,b}}{r_{il}^5} 
 +
 3 \frac{G m_l \delta_{bc}r_{il,a}}{r_{il}^5} 
 -
 15 \frac{G m_l r_{il,a}\,r_{il,b}r_{il,c}}{r_{il}^7}   
 \right)
{\varr}^{(1)}_{i,b,\xi}{\varr}^{(1)}_{i,c,\eta}
\nonumber
\\
&
+ 
\sum_{k,\;k\neq i}
\sum_b
\sum_c
\left(
 -3\frac{Gm_k\,\delta_{ab} r_{ik,c}}{r_{ik}^5}  
 -
 3 \frac{G m_k \delta_{ac} r_{ik,b}}{r_{ik}^5}
 -
 3 \frac{G m_k \delta_{bc} r_{ik,a}}{r_{ik}^5}
 +
 15 \frac{G m_k r_{ik,a}\,r_{ik,b} r_{ik,c}  }{r_{ik}^7}
 \right)
{\varr}^{(1)}_{i,b,\xi}{\varr}^{(1)}_{k,c,\eta}
\nonumber
\\
&
+ 
\sum_{j,\;j\neq i}
\sum_b
\sum_c
\left(
 - 
 3\frac{Gm_j\,\delta_{ab} }{r_{ij}^5} r_{ij,c} 
 -
 3 \frac{G m_j \delta_{ac}}{r_{ij}^5}  r_{ij,b}
 -
 3 \frac{G m_j \delta_{bc}}{r_{ij}^5} r_{ij,a}
 +
 15 \frac{G m_j r_{ij,a}\,r_{ij,b}}{r_{ij}^7} r_{ij,c} 
 \right)
{\varr}^{(1)}_{j,b,\xi}{\varr}^{(1)}_{i,c,\eta}
\nonumber
\\
&
+ 
\sum_{j,\;j\neq i}
\sum_b
\sum_c
\left(
 + 
 3\frac{Gm_j\,\delta_{ab} }{r_{ij}^5} r_{ij,c} 
 +
 3 \frac{G m_j \delta_{ac}}{r_{ij}^5}  r_{ij,b}
 +
 3 \frac{G m_j \delta_{bc}}{r_{ij}^5} r_{ij,a}
 -
 15 \frac{G m_j r_{ij,a}\,r_{ij,b}}{r_{ij}^7} r_{ij,c} 
 \right)
{\varr}^{(1)}_{j,b,\xi}{\varr}^{(1)}_{j,c,\eta}
\nonumber
\\
&
+ 
\sum_b
\sum_{j,\;j\neq i}
\left(
	+\frac{G}{r_{ij}^3} \delta_{ab} 
	-3\frac{G\,{r}_{ij,a} }{r_{ij}^5} r_{ij,b}
	\right)
{\varr}^{(1)}_{ji,b,\xi}{\varm}^{(1)}_{j,\eta}
+ 
\sum_b
\sum_{j,\;j\neq i}
\left(
	+\frac{G}{r_{ij}^3} \delta_{ab} 
	-3\frac{G\,{r}_{ij,a} }{r_{ij}^5} r_{ij,b}
	\right)
{\varr}^{(1)}_{ji,b,\eta}{\varm}^{(1)}_{j,\xi}.
\end{align}

\noindent Note that the first line has the same form as for the first-order variational equations.
This is the linear part of the second-order variational equation. 
The other lines correspond to the non-linear part that couples to the first-order differential equations of the variations $\xi$ and $\eta$.
We can simplify the above expression slightly and convert it to a somewhat more readable vector notation, arriving at
\begin{align}
\dot{\pmb \varv} ^{(2)}_{i,\xi\eta} =& 
\sum_{j,\,j\neq i}
\left(
-\frac{Gm_j\, {\pmb \varr}^{(2)}_{ij,\xi\eta} }{r_{ij}^3} 
+ 3 \frac{G m_j {\pmb r}_{ij}}{r_{ij}^5}
\left(
{\pmb r}_{ij}
\cdot
  {\pmb \varr}^{(2)}_{ij,\xi\eta} 
 \right)
  -
	\frac{G\,{\pmb r}_{ij} }{r_{ij}^3} \varm_{j,\xi\eta}^{(2)} 
 \right)
\nonumber
\\
&
+ 
\sum_{j,\;j\neq i}\left(
 3\frac{Gm_j {\pmb \varr}^{(1)}_{ij,\xi} }{r_{ij}^5} 
 \left(
{\pmb \varr}^{(1)}_{ij,\eta} \cdot {\pmb r}_{ij}
\right)
 +
 3 \frac{G m_j{\pmb \varr}^{(1)}_{ij,\eta}}{r_{ij}^5} 
 \left(
{\pmb \varr}^{(1)}_{ij,\xi} \cdot {\pmb r}_{ij}
\right)
 +
 3 \frac{G m_j {\pmb r}_{ij}}{r_{ij}^5} 
 \left(
{\pmb \varr}^{(1)}_{ij,\xi}\cdot {\pmb \varr}^{(1)}_{ij,\eta}
\right)
 -
 15 \frac{G m_j {\pmb r}_{ij}}{r_{ij}^7}   
 \left(
{\pmb \varr}^{(1)}_{ij,\xi}
\cdot
{\pmb r}_{ij}
\right)
 \left(
{\pmb \varr}^{(1)}_{ij,\eta}
\cdot
{\pmb r}_{ij}
\right)
 \right)
\nonumber
\\
&
+ 
\sum_{j,\;j\neq i}
\left(
-
	\frac{G}{r_{ij}^3}  
{\pmb \varr}^{(1)}_{ij,\xi}{\varm}^{(1)}_{j,\eta}
	+3\frac{G\,{\pmb r}_{ij} }{r_{ij}^5} 
	\left(
	{\pmb r}_{ij} \cdot 
{\pmb \varr}^{(1)}_{ij,\xi}
	\right)
{\varm}^{(1)}_{j,\eta}
	\right)
+ 
\sum_{j,\;j\neq i}
\left(
-
	\frac{G}{r_{ij}^3}  
{\pmb \varr}^{(1)}_{ij,\eta}{\varm}^{(1)}_{j,\xi}
	+3\frac{G\,{\pmb r}_{ij} }{r_{ij}^5} 
	\left(
	{\pmb r}_{ij} \cdot 
{\pmb \varr}^{(1)}_{ij,\eta}
	\right)
{\varm}^{(1)}_{j,\xi}
	\right). \label{eq:vdot2ndorder}
\end{align}
We can use this equation to read off the matrix elements of $X^{(1)}$ and $X^{(2)}$ by comparing the above with Eqs.~\ref{eq:ve1} and \ref{eq:ve2}. 

\end{strip}

\section{Implementation}
\label{sec:implementation}
We have implemented first and second-order variational equations into the $N$-body code \reb \citep{ReinLiu2012}.
\reb is very modular and allows the user to choose from different numerical integrators. 
What we describe here has been tested for the high-accuracy integrator \ias, which is based on a 15th-order Gau\ss-Radau quadrature \citep{ReinSpiegel2015}.
First-order variational equations have also been implemented for the symplectic WHFast integrator \citep{ReinTamayo2015} as a symplectic tangent map \citep{MikkolaInnanen1999}.
In principle, higher-order variational equations could also be implemented as a symplectic tangent map. 
However, the complexity of such a higher-order tangent map goes beyond what we expect to be useful in practice.
We therefore exclusively focus on the general-purpose \ias integrator for the remainder of this paper.

We implement the variational equations in terms of \emph{variational particles}. 
This provides an elegant implementation where variational particles follow a structurally similar (though conceptually different) set of differential equations to the real particles (cf.~Eqs.~\ref{eq:var} and \ref{eq:vdot2ndorder}).
For each first and second order variation that we consider we add $N$ such variational particles to the simulation.
The Cartesian components of a variational particle are then the derivatives of the corresponding real particle's components with respect to the parameter we are varying. 
This implies that the units for different variations will vary (compare with Eq.~\ref{eq:phi-m}).

\subsection{Initialization routines}
In addition to the variational equations themselves, we have implemented convenience methods for initializing the variational particles.
If one is interested in varying one of the cartesian coordinates of a particle, initializing variational particles is as easy as setting all of the coordinates to 0 except one which is set to 1, see Eq.~\ref{eq:initcart}. 
However, as shown above, varying parameters that are non-linear functions of the cartesian coordinates involves calculating first and second derivatives and can quickly become cumbersome.
We are particularly interested in applications involving planetary systems.
We therefore provide routines that allow the initialization of variational particles with respect to changing a particle's mass $m$, as well as its orbit's semi-major axis $a$, eccentricity $e$, inclination $i$, longitude of the ascending node $\Omega$, argument of pericenter $\omega$ and true anomaly $f$.

Since we are doing this to second order for 7 orbital elements\footnote{This includes the mass of the particle.}, we thus have $7+ 7\cdot (7+1)/2=35$ different functions. 
In principle one could also initialize variational particles by calculating finite differences, i.e. creating a second particle with one orbital parameter shifted by a small amount $\alpha$, subtracting each component from the un-shifted particle and then dividing by $\alpha$.
The problem is that this procedure easily leads to numerical issues as the shift $\alpha$ needs to be small enough to be in the linear (quadratic) regime, but large enough to avoid any rounding error due to limited floating point precision.
Our functions that calculate the derivatives analytically avoid this issue.
Our current implementation does not support Jacobi coordinates and assumes that all parameters are given with respect to a fixed central object (heliocentric frame).

We have also implemented a routine that moves the entire system to the centre of mass frame and corrects the variational particles' positions and velocity coordinates consistently.

It is worth pointing out that automatic differentiation (AD) would be well suited to automate this task for even more complicated initialization routines \citep{Neidinger2010}.

\subsection{Test particles}
We call a particle in an $N$-body simulation a test particle if it does not affect other particles in the simulation because it has no mass, $m_i=0$.
If one is interested in the effect of varying the initial conditions of test particles, then the variational equations simplify significantly. 
Because variations of a test particle do not affect the variations of other particles, one can reduce the dimensionality of the first-order variational differential equation, Eq.~\ref{eq:ve1}, from $6N$ to $6$. 
This speeds up the calculation significantly and we have implemented this as an optional flag that can be set when a set of variational equations is initialized.
This might become particularly useful if an approximation of the derivatives is sufficient for a given application.

\subsection{Syntax}
\label{sec:syntax}
Here, we briefly demonstrate how to initialize and run a simulation using the python interface to \reb.
We do this because the layer of abstraction that we came up with to hide the complicated expressions for second order variational equations is essential in making this tool useable in a real world scenario. 
We provide the full documentation for how to use variational equations within \reb online at \url{http://rebound.readthedocs.org}.

A simulation of one planet orbiting a central star can be setup with the following code in \texttt{python}: 
\begin{lstlisting}
import rebound
sim = rebound.Simulation()
sim.add(m=1.)
sim.add(m=0.001, a=1.)
\end{lstlisting}
By default \reb uses units in which $G=1$. One set of first-order variational particles can be set up with a single command:
\begin{lstlisting}
var_i = sim.add_variation()
\end{lstlisting}
The \texttt{var\_i} object contains all the information of this set of variational particles, e.g. the order, the location of variational particles, etc. 
By default, the variational particles' position, velocity and mass coordinates are initialized to zero.
For this example, let us assume that we want to vary the planet's semi-major axis.
We initialise the planet's variational particle using the following command:
\begin{lstlisting}
var_i.vary(1,"a")
\end{lstlisting}
In the background, this command first calculates the orbital parameters of the particle with index 1 (the planet) in heliocentric coordinates. 
Then, the variational particle is initialized using the analytic derivative with respect to the semi-major axis, see Eq.\ref{eq:init1st}.
We can now integrate the system forward in time for, say, 100 time units:
\begin{lstlisting}
sim.integrate(100.)
\end{lstlisting}
The planet's x-position after the integration can be accessed via \texttt{sim.particles[1].x}.
Let us use the result of our integration to estimate the planet's x-position assuming its initial semi-major axis was shifted by $\Delta a = 0.01$.
This can be achieved with the following code
\begin{lstlisting}
Delta_a = 0.01
print sim.particles[1].x 
    + Delta_a * var_i.particles[+1].x
\end{lstlisting}
which should be compared with Eq.~\ref{eq:powerseries}.
To go beyond first order and include second-order variational equations, we setup the second order variational equations (before the integration) with
\begin{lstlisting}
var_ii = sim.add_variation(order=2, 
    first_order=var_i)
\end{lstlisting}
Note that we need to specify the the corresponding first-order variational particles.
This is because second-order variational particles depend on the first-order variational particles and in principle there can be many different first oder variational particles for different parameters.
The initialization of the particle is identical to before
\begin{lstlisting}
var_ii.vary(1,"a")
\end{lstlisting}
The final position can then be estimated by applying Eq.~\ref{eq:powerseries} as before, but now accurate to second order,
\begin{lstlisting}
print sim.particles[1].x 
    + Delta_a * var_i.particles[1].x
    + 0.5*Delta_a*Delta_a * var_ii.particles[1].x
\end{lstlisting}
More complicated and realistic examples are available in the documentation at \url{http://rebound.readthedocs.org}.

\section{Tests}
\label{sec:tests}
In this section we present various tests of our implementation.
These show not only that the implementation is working correctly, but also what second-order variational equations can be used for.
We plan to follow up on several of these ideas in much more detail in future work.

\subsection{Varying one orbital parameter}
\begin{figure}
 \centering \resizebox{0.99\columnwidth}{!}{\includegraphics{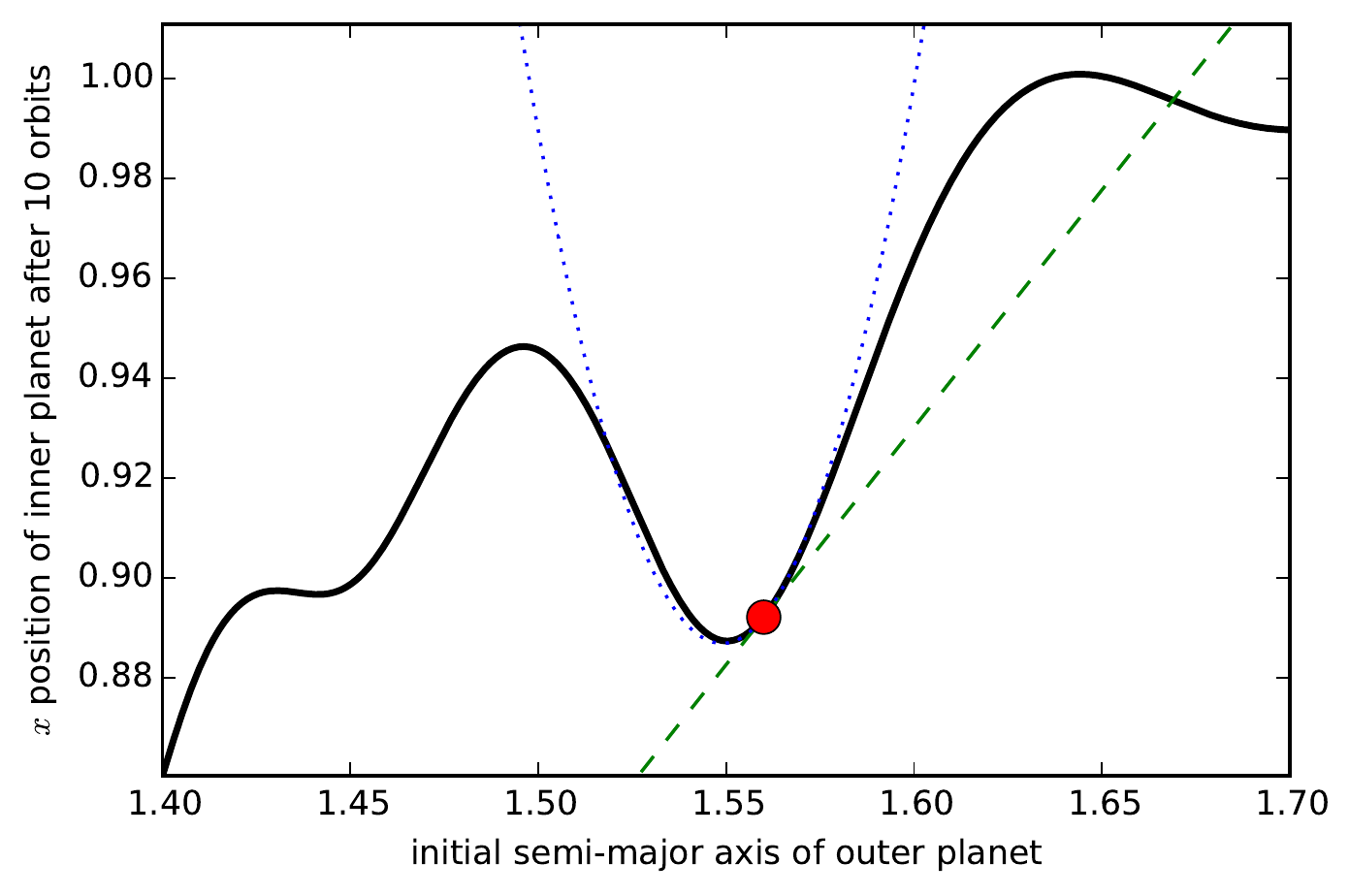}}
 \caption{
     A 10 year integration of a two planet system.
     The black curve shows the the position of the inner planet at the end of the simulation for different initial semi-major axes $a$ of the outer planet.
     Variational equations are integrated for the initial semi-major axis indicated by the red dot.
     Their results are used to approximate the final position of the inner planet as a function of the outer planet's initial semi-major axis $a$.
     The green dashed line uses the first-order variational equations.
     The blue dotted line uses both first and second-order variational equations.
\label{fig:test1}}
\end{figure}

As a first test, we study a two-planet system and vary the initial semi-major axis of the outer planet. 
We use the first and second-order variational equations to approximate the $x$-position of the inner planet after 10 orbits using Eq.~\ref{eq:powerseriesmulti}.
The inner planet's position changes with time because of the planet-star as well as the planet-planet interactions. 

We plot the results in Fig.~\ref{fig:test1}.
The bold black line corresponds to the final $x$-position of the inner planet using a direct $N$-body integration.
The results for both the first and second-order variational equations are shown as a green dashed and blue dotted line, respectively. 
To arrive at these approximations, only one $N$-body simulation with variational equations was run.
Note that this is in contrast to 400 individual $N$-body simulations which were carried out to generate the black curve. 
The red dot indicates the initial semi-major axis used for the single run with variational equations.

As the plot clearly shows, we can use the results from second-order variational equations to accurately predict the final position of the inner planet to within a few percent for initial conditions that are not too far from the original simulation.
Also note that as expected, the second-order variational equations give a significantly better estimate than the first-order equations alone.

\subsection{Optimization problem with one one orbital parameter}
\label{sec:onepar}
\begin{figure}
 \centering \resizebox{0.99\columnwidth}{!}{\includegraphics{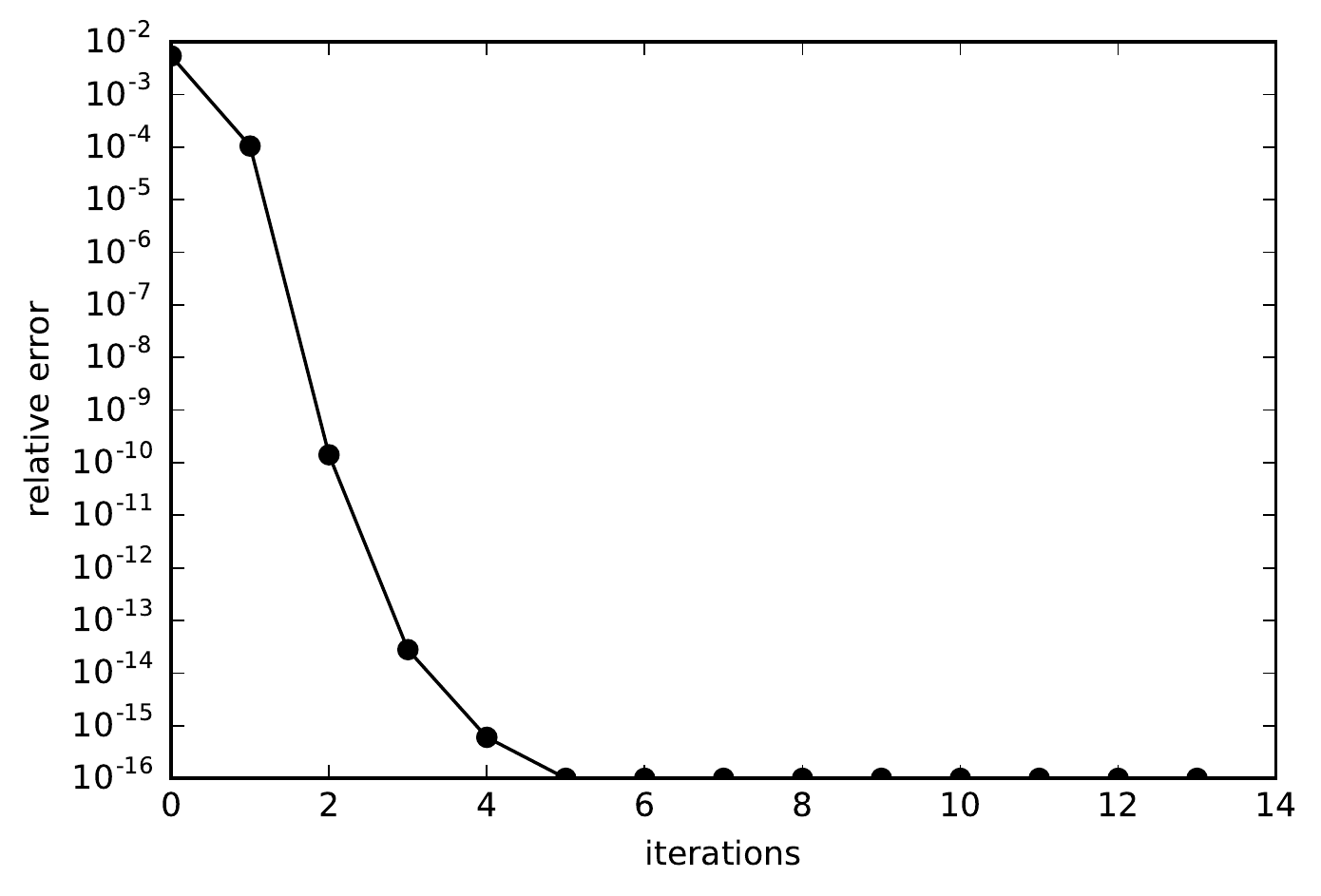}}
 \caption{
     We use first and second-order variational equations in conjunction with Newton's method to solve an optimization problem in a two planet system. 
     The vertical axis shows the relative position offset from the optimum.
     Machine precision is reached after four iterations.
\label{fig:test2}}
\end{figure}
We continue to work with the above two-planet system.
We now attempt to find the initial semi-major axis $a_0$ of the outer planet that minimizes the $x$ coordinate of the inner planet at the end of the simulation. 
This test therefore represents a simple case of a wide range of optimization problems.
Instead of minimizing the $x$ coordinate of the planet, we could also minimize the distance to another planet, or maximize the velocity.
Furthermore, one could replace one of the planets with a spacecraft and then search for an optimal spacecraft trajectory that uses a minimal amount of fuel to reach a final point, and so on.

We use the standard Newton's method to find the optimal value, $x_{\rm min}$.
For that we need the first and second derivatives of the planet's $x$ position (we are looking for the root of the first derivative).
We calculate these using the variational equations.
As a starting point in Newton's method, we use the red dot in~Fig.~\ref{fig:test1}.

We plot the results in Fig.~\ref{fig:test2}. 
The vertical axis shows the relative position offset $\bar x = \left| (x-x_{\rm min})/{x_{\rm min}}\right|$ as a function of the iteration.
After four iterations, the method has converged to machine precision.
With any derivative free method such as the bisection method we would need more iterations to achieve machine precision.

\subsection{Fitting a radial velocity curve}
\label{sec:rvtest}
\begin{figure}
 \centering \resizebox{0.99\columnwidth}{!}{\includegraphics{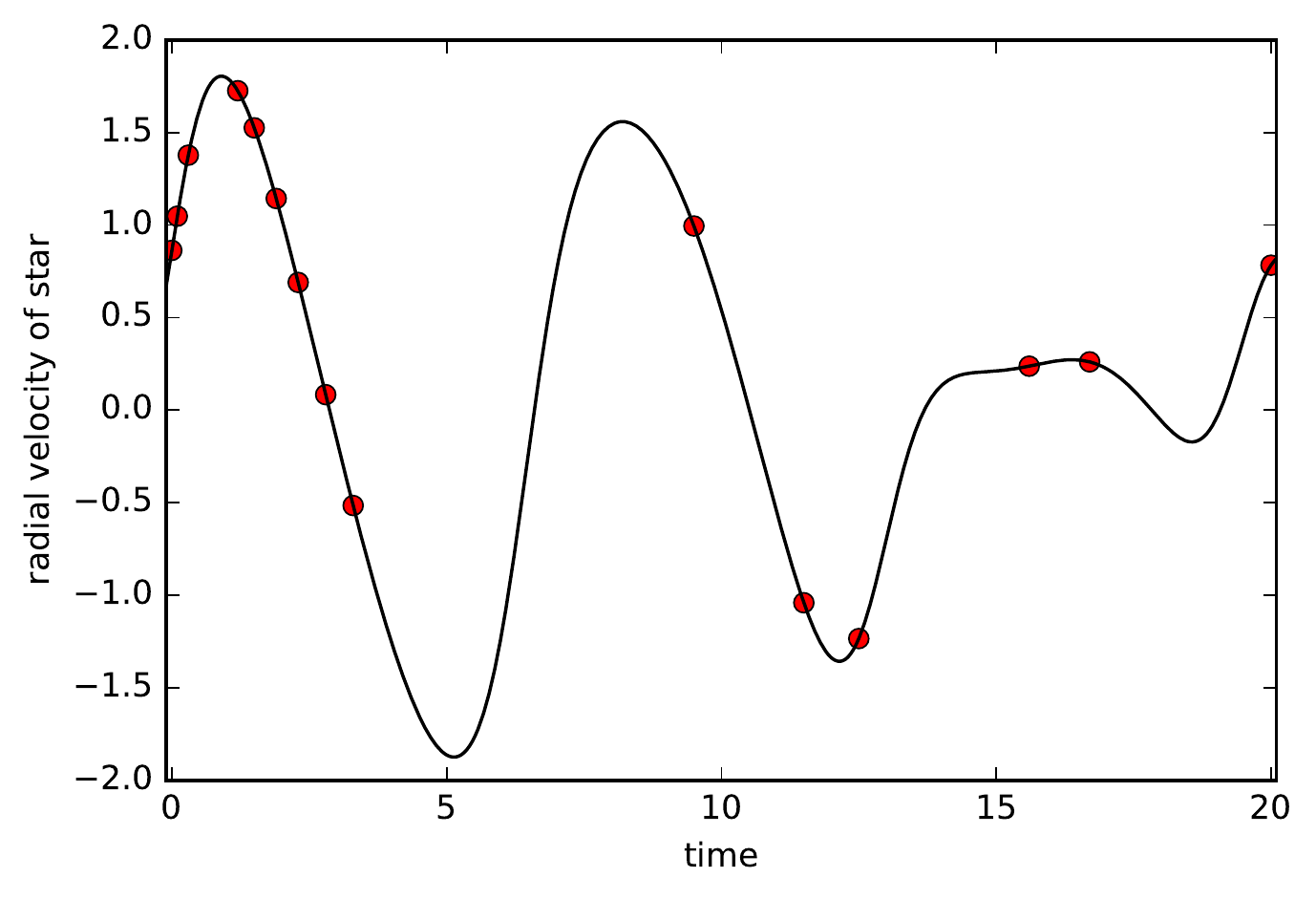}}
 \caption{
     A synthetic radial velocity curve of a two planet system.
     The red dots indicate where datapoints were taken.
    \label{fig:test3a}}
\end{figure}
\begin{figure}
 \centering \resizebox{0.99\columnwidth}{!}{\includegraphics{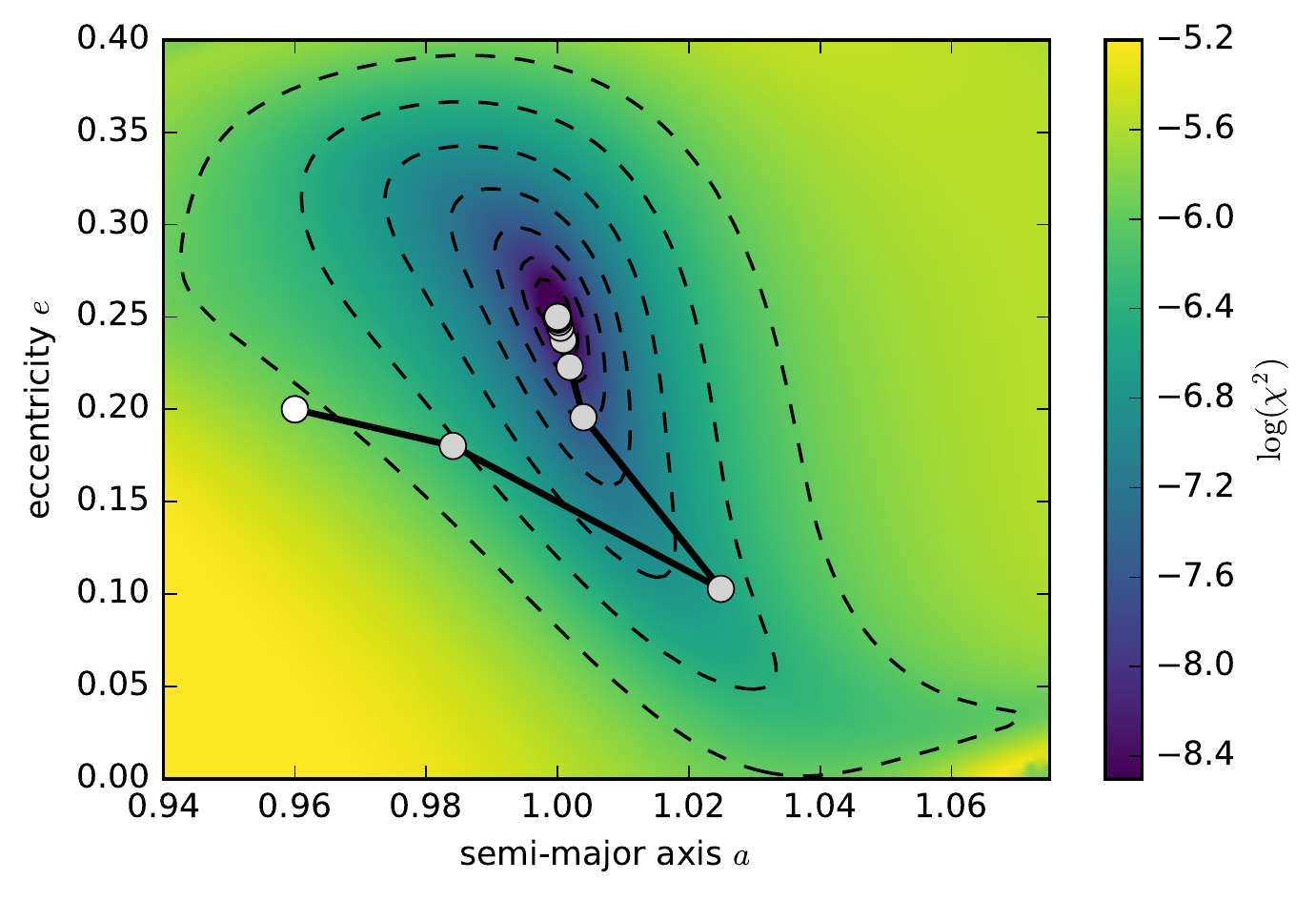}}
 \caption{
     Minimizing $\chi^2$ to fit the radial velocity curve in Fig.~\ref{fig:test3a}. 
     Second-order variational equations, Newton's method and the softabs metric are used to descend to the minimum.
    \label{fig:test3b}}
\end{figure}
We now present a more complicated example in which we attempt to fit the reflex motion of a star in a two-planet system to a synthetic radial velocity data set. 
In Fig.~\ref{fig:test3a} we show the synthetic radial velocity curve of the star as a function of time (the units are irrelevant for this discussion).
The red dots show where an observation is taken. 
For simplicity we only vary two orbital parameters, the semi-major axis $a$ and the eccentricity $e$ of the inner planet.
All other parameters are the same as in the reference simulation.
Our goal is to match these synthetic observations and thus to find the true parameters $(a, e)$, starting from an arbitrary initial guess of semi-major axis and eccentricity,~$(a_0, e_0)$.

We label the synthetic observation at time $t_i$ with $o_i$ and the radial velocity in our simulation with $v_i$.
The problem can be expressed again as an optimization problem by defining a goodness of fit, e.g., 
\begin{align}
\chi^2 = \sum_i \left(v_i-o_i\right)^2,
\end{align}
which we try to minimize.
Note that $v_i$ are functions of the initial $a$ and $e$. 
More complicated and realistic $\chi^2$ functions that take into account observational uncertainties can be easily constructed, but we here work with the simplest case.
The chain rule yields
\begin{align}
\frac{\partial}{\partial a}\chi^2 &= 2 \sum_i \left(v_i-o_i\right) \frac{\partial v_i}{\partial a}\\
\frac{\partial^2}{\partial a^2}\chi^2 &= 2 \sum_i\left( \left(v_i-o_i\right) \frac{\partial^2 v_i}{\partial a^2}+ \left(\frac{\partial v_i}{\partial a}\right)^2\right)
\end{align}
and similar expressions for the derivatives with respect to $e$ and the cross-term~$\nicefrac{\partial^2 \chi^2}{\partial a \partial e}$.
The second-order variational equations are used to calculate the derivatives involving $v_i$.
We can then use the standard Newton's method to iterate and find the extremum in $\chi^2$:
\begin{align}
\begin{pmatrix}
a_{n+1}\\
e_{n+1}
\end{pmatrix}
=
\begin{pmatrix}
a_n\\
e_n
\end{pmatrix}
-
\begin{pmatrix}
\frac{\partial^2 \chi^2}{\partial a^2} & 
\frac{\partial^2 \chi^2}{\partial a\partial e} \\ 
\frac{\partial^2 \chi^2}{\partial a\partial e}& 
\frac{\partial^2 \chi^2}{\partial e^2} 
\end{pmatrix}^{-1}
\cdot 
\begin{pmatrix}
\frac{\partial \chi^2}{\partial a} \\ 
\frac{\partial \chi^2}{\partial e} 
\end{pmatrix}.
\end{align}
The matrix in the above equation is the inverse of the Hessian of $\chi^2$, $H^{-1}$.
Newton's method will only converge where $\chi^2$ is convex, or in other words where the matrix $H$ is positive definite.
To increase the convergence region we use a trick to ensure that $H$ is positive definite everywhere by using the softabs metric of the Hessian $\sabs{H}$, instead of $H$ itself \citep{Betancourt2013}.
The modified Newton's method becomes
\begin{align}
\begin{pmatrix}
a_{n+1}\\
e_{n+1}
\end{pmatrix}
=
\begin{pmatrix}
a_n\\
e_n
\end{pmatrix}
-
\sabs{H}^{-1}
\cdot 
\begin{pmatrix}
\frac{\partial \chi^2}{\partial a} \\ 
\frac{\partial \chi^2}{\partial e} 
\end{pmatrix}.
\end{align}

In Fig.~\ref{fig:test3b} we plot the relevant part of the parameter space.
The colours and contours correspond to the logarithm of $\chi^2$.
We start the iteration at $(a_0,e_0) = (0.96,0.2)$ and converge to the true minimum within machine precision in less than ten iterations. 
Newton's method converges to the global minimum for most nearby starting values (those near the centre in Fig.~\ref{fig:test3b}). 
If the initial conditions are far from the global optimum, then, as expected, the method might not converge to the global minimum.
We note that this problem of non-convergence is a feature of the adopted optimization algorithm (Newton's method), and not of the variational equations.
In particular, even in cases where Newton's method does not converge, the derivatives are calculated exactly (to machine precision).

For the above reasons, the method presented in this example is not well suited for finding the global minimum within a complex parameter space.
Other methods such as simulated annealing or parallel tempering are most likely faster and more reliable. 
However, as we discuss below, a combination of methods is a promising future area of research if one can make use the second-order variational equations to converge to a local optimum within almost constant time (or $\mathcal{O}(1)$ iterations)\footnote{Newton's method converges quadratically, i.e. the number of significant digits roughly doubles after every iteration. Thus, if we are close to a local minimum and we work in double floating point precision with 16 significant digits, we need $\sim 4$~iterations to converge to machine precision.}.

\subsection{Comparison to a finite difference approach}
\label{sec:finitedifference}
\begin{figure*}
 \centering \resizebox{0.99\textwidth}{!}{\includegraphics{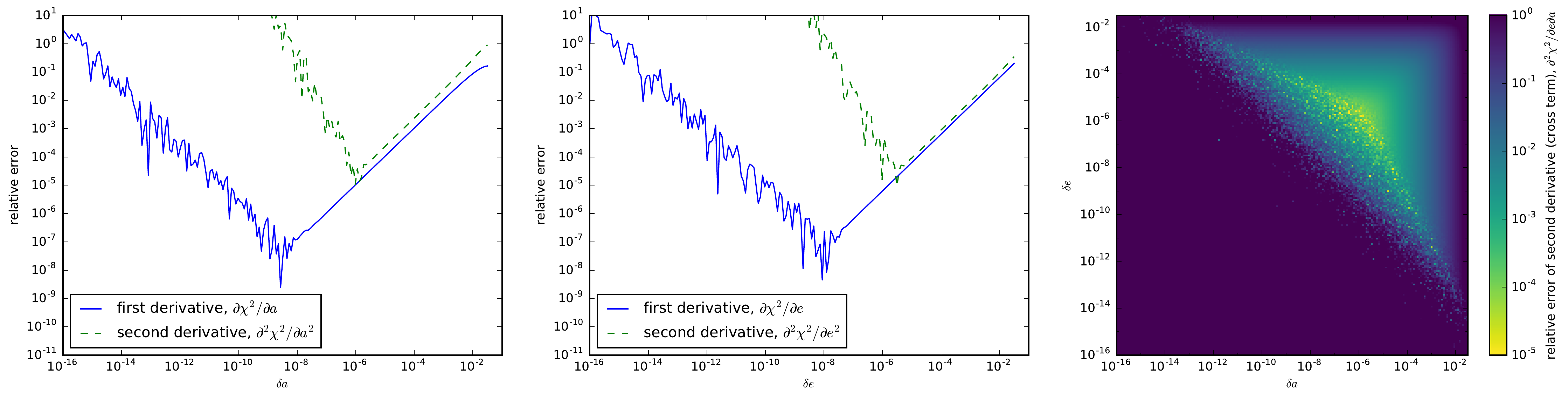}}
 \caption{
     Relative error in calculating derivatives using the finite difference approach in the optimization problem described in Sec.~\ref{sec:rvtest}. 
     The plots show the relative error of first and second derivatives of $\chi^2$ as a function the initial finite difference $\delta a$ and $\delta e$. 
     Using variational equations, this problem does not exist and derivatives are exact up to machine precision.
    \label{fig:test4}}
\end{figure*}

In the above optimization problem, we use variational equations to calculate the first and second derivatives of $\chi^2$. 
One can also use a finite difference approach to estimate the derivatives. 
As we show in this section, this is not viable in most scenarios as two separate competing constraints require fine tuning of the finite difference parameters.

Let us try to calculate all the first and second order derivatives that we need in the radial velocity fit problem from Sec.~\ref{sec:rvtest}: $\nicefrac{\partial \chi^2}{\partial a}$, $\nicefrac{\partial \chi^2}{\partial e}$, $\nicefrac{\partial^2 \chi^2}{\partial a^2}$, $\nicefrac{\partial^2 \chi^2}{\partial e^2}$ and the cross term $\nicefrac{\partial^2 \chi^2}{\partial a \partial e}$.
To use the finite difference method, we need to choose a \emph{finite} initial differences $\delta a$ and $\delta e$. 
These are then used to initialize the orbits of shadow particles which are integrated using the normal equations of motion.

The actual value of $\delta a$ and $\delta e$ is crucial.
It has to be small enough to ensure the simulation remains in a linear regime (quadratic for second order). 
However, making the finite differences too small results in loss of accuracy due to finite floating point precision. 
Thus there is an optimum between these two competing effects.
The precise value is problem specific.
For the radial velocity test case we find an optimum around $\delta a \sim 10^{-8}$ for first-order derivatives and $\delta a \sim 10^{-6}$ for second-order derivatives.

This problem is illustrated in Fig.~\ref{fig:test4}. 
We plot the relative error of the first and second-order derivative as a function of the initial finite differences $\delta a$ and $\delta e$. 
One can see that the best possible estimate of the first derivative is only accurate to within $10^{-7}$. 
Worse yet, the best estimate of the second derivative is only accurate to within $10^{-4}$. 
Using the finite difference approach we cannot obtain a better estimate.

The problem gets even worse if one is interested in the cross-term in the Jacobian, e.g.  $\nicefrac{\partial^2 \chi^2}{\partial a \partial e}$.
The relative error of this quantity is plotted in the right panel of Fig.~\ref{fig:test4}. 
The cross term depends on both finite differences $\delta a$ and $\delta e$. 
There is only a small area in the $\delta a$/$\delta e$ space that gives reasonably accurate results. 
Finding the best combination of the initial finite differences is difficult, requires problem-specific fine tuning and becomes quickly infeasible, especially for applications where a wide range of the physical parameter space is explored.

None of these problems exist using the variational equation approach that we present in this paper.
Note that we also do not use finite difference for the initialization of variational particles.
The entire framework does not contain any small parameters that could lead to numerical problems (cf. $\delta a, \delta e$). 

It is worth pointing out that the \ias integrator that we use for the normal $N$-body integration as well as for the variational equations is accurate to machine precision \citep{ReinSpiegel2015}.  
Furthermore, the energy error in long-term simulations grows sub-linearly and follows Brouwer's Law \citep{Newcomb1899,Brouwer1937}.
\ias is therefore as exact as any integrator can possibly be\footnote{To within a constant factor of a few.}, using only double precision arithmatic. 
This statement also applies to the variational equations and therefor to the derivatives we calculate with their help.
All derivatives are exact up to machine precision.
They can not be calculated more accurately without going to extended precision.

\subsection{Runtime}
\label{sec:rvtest}
\begin{figure}
 \centering \resizebox{0.99\columnwidth}{!}{\includegraphics{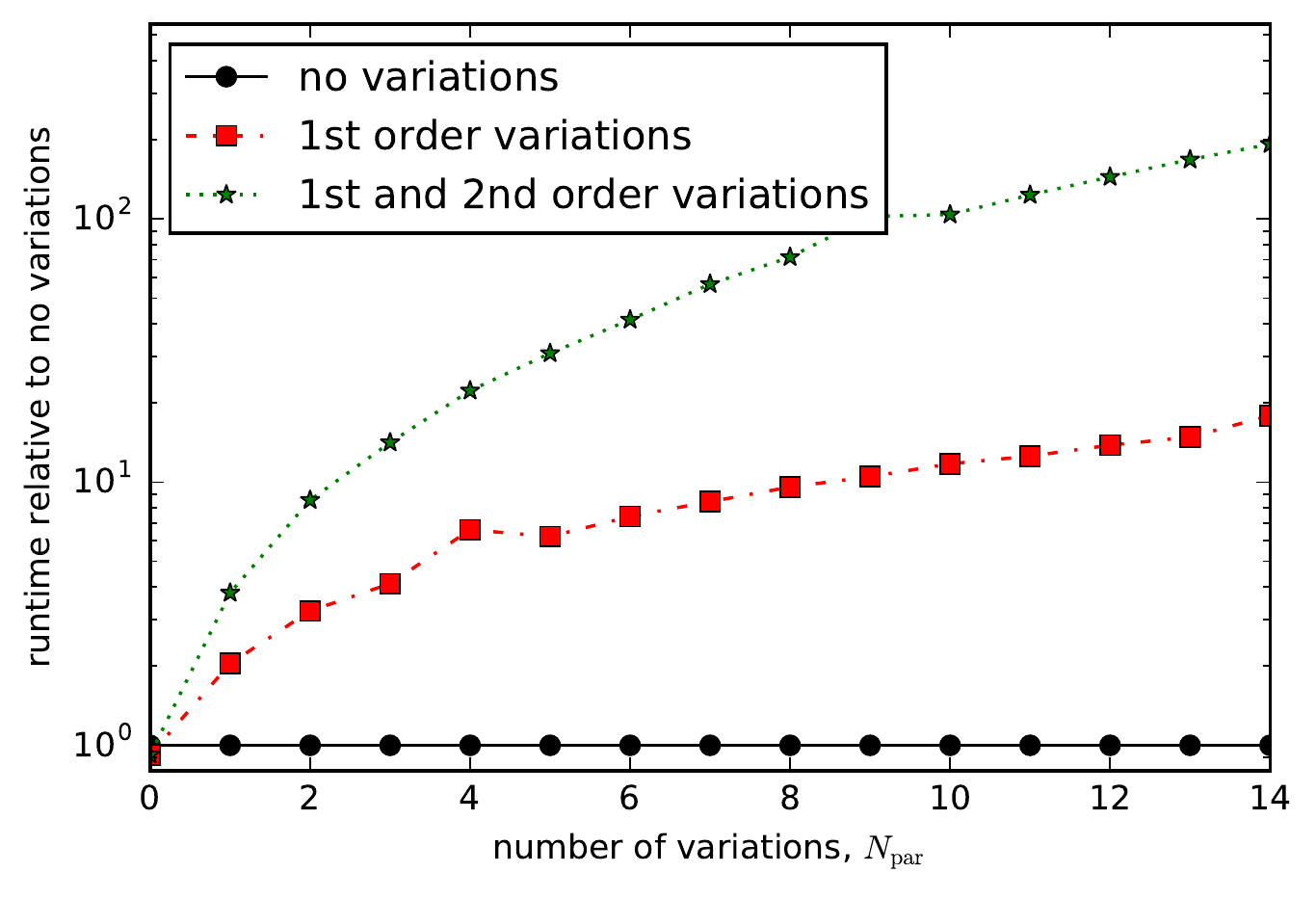}}
 \caption{
     Runtime of a simulation with variational equations of first and second order relative to a simulation without variational equations. 
    \label{fig:test5}}
\end{figure}
One thing to keep in mind for optimization problems is the computational complexity of a simulation with first and second-order variational equations.
If there are $N$ particles and $N_{\rm par}$ free parameters, then the computation time for a simulation with second-order variational equations scales as $N^2(1+N_{\rm par} + \frac12 N_{\rm par} (N_{\rm par}+1)$. 

We tested this scaling in a simulation of two planets in which we vary all 14~planet parameters (all orbital parameters and the masses). 
The results are plotted in Fig.~\ref{fig:test5} and agree with our estimate.

If every parameter of every particle is varied, one ends up with a runtime that scales approximately as $\frac12 N^4$. 
This indicates that using variational equations might only be competitive when combined with other methods.
However, if another method brings us close to a local minimum (using, e.g., simulated annealing, parallel tempering), then an approach based on variational equations can converge to the local optimum within just a few iterations, in (almost) constant time.

\section{Conclusions}
\label{sec:conclusions}
In this paper we presented the theoretical framework for using second-order variational equations in $N$-body simulations to estimate how particle trajectories vary with respect to their initial conditions.
We described a flexible implementation of these equations within the \reb integrator package.

A major motivation for developing first-order variational equations was to overcome the numerical inaccuracies associated with finite-difference methods that use shadow particles \citep[e.g.,][]{Tancredi2001}.
We showed in Sec.~\ref{sec:finitedifference} that this problem is exacerbated at second order, requiring careful problem-dependent fine tuning.
Additionally, the number of shadow particles required by the finite-difference approach is always the same as the corresponding number of variational equations to follow.
The variational approach is therefore much more robust and effectively equal in speed.  

An important application for second-order variational equations is in solving optimization problems.
First derivatives furnish only the right {\it direction} to move in a parameter landscape toward a minimum; second derivatives provide a {\it scale} for how far one must jump to reach that minimum.
If near a minimum, the first and second derivatives furnished by the variational equations can converge to within machine precision of the minimum in just a few iterations.
We illustrated this behaviour in both a simple two-planet case (Sec.~\ref{sec:onepar}) and in the fitting of a radial velocity curve (Sec.~\ref{sec:rvtest}).
Variational equations might also be applied to spacecraft trajectory optimization or asteroid deflection.

The optimization problem presented in this paper uses second order variational equations in connection with the classical optimization algorithm of Newton.
One can also use the second-order variational equations in connection with a Markov Chain Monte Carlo (MCMC) method.
Specifically, for both Riemann Manifold Langevin and Hamiltonian Monte Carlo methods, higher order derivatives, and therefore higher order variational equations, are essential \citep{GirolamiCalderhead2011}. 
A full discussion of these MCMC methods and their application goes beyond the scope of this paper but we note that our initial tests of these methods show great promise.
In particular, we observe very short auto-correlation times when using a Riemann Manifold Langevin MCMC to sample the posterior of radial velocity curves.

For a long time, first-order variational equations have been widely used to calculate Lyapunov exponents and the Mean Exponential Growth of Nearby Orbits (MEGNO, \citealt{Cincotta2003}) in the astrophysics community \citep{Tancredi2001,Hinse2010}.  
We speculate that higher-order variational equations may be able to improve such chaos indicators.
Since only one set of variational equations is needed for the calculation of the Lyapunov exponent, including second-order variational equations will keep the numerical scaling $\mathcal{O}(N^2)$ and only increase the computational cost by 50\%.

The latest version of \reb includes the second-order variational equations and can be downloaded at \url{https://github/com/hannorein/rebound}.
The package is free to use under an open source license.
We provide also a \texttt{git} repository with \texttt{jupiter} notebooks to reproduce the figures in this paper at \url{https://github.com/hannorein/variations}.
The notebooks can be run interactively in the web browser without the need to install any software locally.

\section*{Acknowledgments}
We thank Eric Ford, Benjamin Nelson and Scott Tremaine for many helpful discussions and Gottfried Wilhelm Leibniz for the chain rule.
This research has been supported by the NSERC Discovery Grant RGPIN-2014-04553. 
D.T. is grateful for support from the Jeffrey L. Bishop Fellowship.
This research made use of \texttt{iPython} \citep{ipython}, \texttt{SciPy} \citep{scipy} and \texttt{matplotlib} \citep{Hunter2007}.

\bibliography{full}

\end{document}